\documentclass{elsart}
\usepackage{amssymb}
\usepackage{amsmath}
\usepackage[dvips]{graphicx}
\journal{International Journal of Non-Linear Mechanics\quad}
\input{tcilatex}

\begin{document}

\begin{frontmatter}

\title
{Properties of thermocapillary fluids \\ and symmetrization of motion equations}

\author{Henri Gouin}
\ead{henri.gouin@univ-amu.fr;henri.gouin@yahoo.fr}

\address {Aix-Marseille Univ, CNRS, Centrale Marseille,
 M2P2 UMR 7340,\\
13451 Marseille, France}

\address {}

\begin{abstract}
The equations of fluid motions are considered  in the case of internal energy depending on mass density,  volume entropy and their spatial derivatives. The model corresponds to domains with large density gradients   in which the temperature is not necessary uniform. In this new general representation writes in symmetric form with respect to the mass and entropy densities.     For conservative motions of perfect \emph{thermocapillary fluids}, Kelvin's circulation theorems are always valid. Dissipative cases are also considered; we
obtain the balance of energy and  we prove that  equations
are compatible with the second law of thermodynamics.
 The  internal energy form  allows to obtain  a Legendre transformation inducing a quasi-linear system of conservation laws  which can be written in a divergence form and the stability  near equilibrium positions can be deduced. The result extends  classical hyperbolicity theory   for governing-equations' systems in hydrodynamics, but   symmetric matrices are replaced by Hermitian matrices.

\end{abstract}


\end{frontmatter}

\section{\label{sec1}Introduction}

 Theoretical and experimental
studies show that, when working far from critical conditions, the liquid-vapour
capillary layer  has a few molecular-beams' thickness \cite%
{Hamak,Ono,Israel}. Consequently, liquid-vapour interfaces are generally represented by
material surfaces endowed with  surface energy related to Laplace's surface-tension \cite{Gatignol}.  The surfaces have their own characteristic
behaviours and energy properties \cite{Levitch}.
 In
interfacial layers, molecular models  -  as used in kinetic theory of gases - express   behaviours associated with non-convex internal
energies \cite{Cahn,Hohenberg,Widom,rowlinson}.
These models appear advantageous as they provide a  more precise
verification of Maxwell's rule applied to isothermal phase-transitions \cite%
{Widom,Gouin4}. Nonetheless, they present two disadvantages.
First, for densities that lie between bulk densities, the pressure may
become negative. However, simple physical experiments can be used  to cause
traction that leads to  negative pressure values \cite{Rocard}.
Second, in the field between bulks, internal energies cannot be represented by
 convex surfaces associated with the variation of densities.
The fact seems to contradict the existence of equilibrium states.
To overcome the  disadvantages, the thermodynamics replaces
the non-convex portions corresponding to internal energies by  planar domains \cite{callen};
the fluid can no longer be considered as a continuous medium.

At equilibrium, an appropriate modification of the layer stress-tensor, expressed in an anisotropic form, can eliminate the previous disadvantages;
then, the continuous-medium energies change   \cite{Cahn,rowlinson} and
 near the critical point,   allow  to
study  interfaces of non-molecular size \cite{Evans,domb}. The approach is not new and
 dates back to van der Walls \cite{Waals,Widom2} and Korteweg \cite{Korteweg};
it corresponds to what is known as a Landau-Ginzburg theory \cite%
{Hohenberg}. The  contradiction between
Korteweg's classical stress theory and the Clausius-Duhem inequality \cite%
{Gurtin} is corrected by Eglit \cite{Eglit}, Dunn and Serrin \cite{Dunn},
Casal and Gouin \cite{Gouin5}.
\newline
To study capillary layers and bulks, the \emph{second gradient theory} \cite{Germain,Germain1} - conceptually simpler than   Laplace's theory -  led to a capillary model for isothermal liquid-vapour interfaces. Fluids endowed with internal capillarity yield equations of motion and energy  including additive terms. The internal energy of such fluids is a function of the entropy, the mass density and the gradient of mass density \cite{Gouin1,Biguenet,Gouin2}. Gradient theory can be extended to solid mechanics,  materials, nanofluidics, fluid mixtures \cite{GavrilyukS,Eremeyev,Dell'Isola1,Forest,Saccomandi,Gouin7} and   developed at $n$-order ($n>2$)   \cite{Gouin3}.\newline
The simplest model in
continuum mechanics considers a volume internal energy $\varepsilon$ as
the sum of two terms: a first one $\varepsilon_{_0}$ corresponding to a medium with
uniform composition equal to the local one and a second one
associated with the non-uniformity of the fluid  and is
approximated by a gradient expansion, typically truncated to the
second order \cite{Cahn,Waals}:
$$
\varepsilon =\varepsilon_{_0} (\rho ,\eta)+{\frac{1}{2}}%
\,m\,(\func{grad}\rho )^{2},
$$
where $\rho $ is the mass
density (or volume mass), $\eta$ the volume entropy, $\varepsilon_{_0}$ the volume internal-energy
of the fluid assumed  to be homogeneous and $m$ is a coefficient independent of $\eta$, $\func{grad}\rho $ and of any higher derivatives \cite{rowlinson}. In such a model, $\eta$
varies with $\rho $ through   isothermal interface in the same way as in
bulks  and at
given temperature $T$ satisfies
\begin{equation}
\frac{\partial \varepsilon_{_0} }{\partial \eta}(\rho ,\eta)=T , \label{temp0}
\end{equation}
so,\ \ $\varepsilon_{_0} =\varepsilon_{_0}
(\rho, T)$. At given temperature $T$, the points representing  phase states in the
$(\rho , \eta, \varepsilon_{_0})$ space lie on a curve instead of    surface
$\varepsilon_{_0} =\varepsilon_{_0} (\rho ,\eta)$. In fact, the assumption is not exact
for  realistic potentials;
in practice the potential for
the two-density form of  van der Waals' theory is not
constructed by   prescription (\ref{temp0}) but by other means  \cite{rowlinson} (Ch. 8). Aside
from the question of accuracy, there are   qualitative features
of some interfaces, especially in systems of
more than one component, that require two or more independently
varying densities - entropy included - for their description; in fact, when we  have non-monotonic behaviours, one-density models inevitably lead to  monotonic variations of densities  \cite{Widom}.
In our case, the model must be
extended by taking account of not only the strong variations of
matter density through   interfacial layers but also the strong
variations of entropy. For this purpose, Rowlinson and Widom in \cite{rowlinson}  (Ch. 3 and Ch. 9)  introduced  an energy
arising from the mean-field theory and depending on the deviations
of   densities $\rho$ and $\eta $  from their values at the
critical point and on the gradients of   densities.    Consequently, we can also imagine non-isothermal steady motions in zones with large density gradients \cite{Sone}.\\

The paper is presented as follows :\newline
In Section 2, we consider different forms of   equation of motions in the most general case.
The Hamilton principle yields the equation  of conservative motions in a symmetric form with respect to    mass and entropy volumes. In \cite{casal} we considered thermocapillary fluids as fluids with a specific internal energy in the form $\alpha = \alpha(\rho, s, \func{grad\, \rho}, \func{grad \,s})$ where $s$ is the specific entropy.
But, it is more convenient to consider the volume entropy in place of the specific entropy to obtain a  simpler system of equations.
\\
In Section 3, we extend the balance equations to   viscous fluids. The equation of energy is completed with a heat flux and a  heat supply. We get an additive \emph{interstitial-working term} similar to  a  heat-flux vector and the processes' equations are compatible with the second law of thermodynamics.\\
In Section 4, we revisit  Kelvin's circulation-theorem and analyse the surface tension of planar interfaces at equilibrium. The Maxwell rule is extended for thermocapillary fluids.\\
Section 5 is a  completely new study. A Legendre transformation yields a system of equations in a divergence form when conjugated variables - with respect to the mass density, volume entropy and their gradients - are used.  The hyperbolicity of the system of governing equations can be studied. Small perturbations near an equilibrium position are analysed. Eigenvalues associated with Hermitian matrices conclude to the stability of equilibrium positions by extending Godunov and Lax-Friedrichs analyses \cite{Godunov,Friedrichs}.\\
A conclusion ends the paper.

For any vectors $\boldsymbol{a,b}$ we use the notation $%
\boldsymbol{a}^{\star }\boldsymbol{b}$ for the scalar product (the
line vector is
multiplied by the column vector) and $\boldsymbol{a}^{\ }\boldsymbol{b}%
^{\star }$ for the tensor product (or $\boldsymbol{a}\otimes
\boldsymbol{b}$ the column vector is multiplied by the line vector),
where  superscript $^\star$ denotes the transposition. Divergence of
a linear transformation $\boldsymbol{D}$ is the covector
$\mathop{\rm div}\boldsymbol{D}$ such that, for any
constant vector $\boldsymbol{c}$, $\ (\mathop{\rm div} \boldsymbol{D})\, \boldsymbol{c}= %
\mathop{\rm div}(\boldsymbol{D}\,\boldsymbol{c})$. The identical transformation is
denoted by $\boldsymbol{I}$.

\section{\label{sec2}Equation of motions}

\subsection{\label{subsec2.1}Preliminaries}

The volume internal energy of a thermocapillary fluid is represented
 by a development in \emph{gradients} with respect to $\rho$ and $\eta$ :
\begin{equation}
\varepsilon = \varepsilon (\rho, \eta, \func{grad} \rho, \func{grad} \eta) .
\label{volumeenergy}
\end{equation}
A particular case of volume internal energy can be
\begin{equation}
\varepsilon = \varepsilon_{_0} (\rho, \eta) + \frac{1}{2}\left( C\, ({\func{%
grad}\rho})^2 + 2\, D\, ({\func{grad} \rho})^\star \func{grad} \eta + E\, ({%
\func{grad}\eta})^2\right) ,  \label{specialvolumeenergy}
\end{equation}
where $C, D, E$ are assumed to be constant; in special case  $D
=0$ and $E= 0$, we get  Cahn and Hilliard's fluids \cite{Cahn}.

$\quad\bullet$ Thermodynamical potential $\varepsilon_{_0} (\rho, \eta) =
\rho\,\alpha(\rho,s)$ is the volume internal energy of the fluid bulk   with volume
mass $\rho$ and volume entropy $\eta$ (the same potential expression as for compressible fluids). Consequently,
\begin{equation*}
d\alpha (\rho,s) = \frac{ P}{\rho^2}\, d\rho+{T}\, ds\,,
\end{equation*}
where $ P$ is the thermodynamical pressure and $ T$   the Kelvin temperature. Then,
\begin{equation*}
d\varepsilon_{_0}= {\mu_{_0}}\, d\rho+{ T}\, d\eta\,,
\end{equation*}
where $\displaystyle {\mu_{_0}} = \frac{\partial\varepsilon_{_0} (\rho,\eta)}{\partial\rho} \,$ is the bulk chemical-potential.
 We get
\begin{equation*}
 P  = \rho\,{\mu_{_0}}+\eta\,{T} -\varepsilon_{_0}
\end{equation*}
and
\begin{equation*}
\label{entropy} \eta\,dT = dP -\rho\, d\mu_{_0}.
\end{equation*}
$\quad\bullet$  For  thermocapillary fluids associated with Eq. (\ref{volumeenergy}) we denote,
\begin{equation}
\begin{array}{l}
\qquad\qquad\qquad d\varepsilon= \mu\,d\rho+ {\mathcal T}\, d\eta + \boldsymbol{\Phi}^\star d(\func{grad}%
\rho) + \boldsymbol{\Psi}^\star d(\func{grad}\eta) \label{diffenergy} \\
\displaystyle \mathrm{with}\quad \mu = \frac{\partial\varepsilon}{\partial\rho}\,, \quad {\mathcal T}  = \frac{\partial\varepsilon}{\partial\eta}\,,\quad \boldsymbol{\Phi}^\star = \frac{%
\partial\varepsilon}{\partial\func{grad}\rho} \,,\quad
\boldsymbol{\Psi}^\star = \frac{\partial\varepsilon}{\partial\func{grad}\eta}\, .%
\end{array}%
\end{equation}
We always denote
\begin{equation*}
\mathcal{P} = \rho\,\mu+\eta\,{\mathcal T} -\varepsilon ,  \label{Pressure}
\end{equation*}
where  $\mathcal{P}$ is called the \emph{thermocapillary pressure},   $\mu$ and ${\mathcal T}$ are extended by
Eq. (\ref{diffenergy}) as the \emph{thermocapillary chemical-potential} and the \emph{thermocapillary temperature}, respectively.

In  the particular case of Eq. (\ref{specialvolumeenergy}) we obtain,
\begin{equation*}
\begin{array}{l}
\displaystyle \mu = \frac{\partial\varepsilon_{_0}}{\partial\rho} \,,
\quad {\mathcal T}  = \frac{\partial\varepsilon_{_0}}{\partial\eta}\, ,\\
\boldsymbol{\Phi}^\star = C ({\func{grad}\rho})^\star + D ({\func{grad}\eta})^\star\quad%
\mathrm{and}\quad\boldsymbol{\Psi}^\star = D ({\func{grad}\rho})^\star + E ({\func{%
grad}\eta})^\star \,,
\end{array}%
\end{equation*}
where $\mu\equiv\mu_{_0}$ and ${\mathcal T}\equiv T$ are also the chemical
 potential and the temperature of  bulks.

\subsection{\label{subsec2.2}The Hamilton principle \cite{Berdichevsky}}

The mass conservation  writes :
\begin{equation}
\frac{\partial \rho }{\partial t}+\func{div}\rho \boldsymbol{u}=0.\label{mass conservation}
\end{equation}%
For isentropic motions, the volume entropy conservation   writes :
\begin{equation}
\frac{\partial \eta }{\partial t}+\func{div}\eta \boldsymbol{u}=0. \label{entropy conservation}
\end{equation}%
The Hamilton action  between time $t_1$ and time $t_2$  is \cite{Lin,Herivel},
\begin{equation*}
S =\int_{t_{1}}^{t_{2}}\int_{D_{t}}L\,dv\,dt\quad \mathrm{with}\quad L=\frac{1%
}{2}\,\rho \,\boldsymbol{u}^{\star }\boldsymbol{u}-\varepsilon -\rho
\,\Omega \,.
\end{equation*}%
where $L$ is the Lagrangian, $dv$ is the volume element of physical space $%
D_{t}$ at time $t$, $dt$ is the time differential, $\boldsymbol{u}$ is the fluid
velocity-vector  and $\Omega $ the external-force potential. We have
the properties associated with the variations of $\boldsymbol{u},\rho $ and $%
\eta $
\begin{equation}
\delta \boldsymbol{u}=\frac{d\boldsymbol{\zeta }}{dt},\quad \delta \rho
=-\rho \,\func{div}\boldsymbol{\zeta },\quad \delta \eta =-\eta \,\func{div}%
\boldsymbol{\zeta },  \label{variations}
\end{equation}%
where $\boldsymbol{\zeta }=\delta \boldsymbol{x}$ notes the variation of
Euler position $\boldsymbol{x}$ as defined by Serrin in \cite{Serrin}. Equation (\ref{variations}$^{3}$)
corresponds to an isentropic variation when the motion is conservative and isentropic.\newline
Thanks to Eqs. (\ref{diffenergy}-\ref{variations}), the variation of Hamilton's
action  is \cite{Seliger},
\begin{eqnarray*}
\delta S &=&\int_{t_{1}}^{t_{2}}\int_{D_{t}}\left[ \rho \,\boldsymbol{u}%
^{\star }\frac{d\boldsymbol{\zeta }}{dt}+(\rho \,\mu -\varepsilon +\eta
\,{\mathcal T} )\func{div}\boldsymbol{\zeta }\right. \, \\
&&\left. -\delta ({\func{grad}\rho })^{\star }\,\boldsymbol{\Phi }-\delta ({%
\func{grad}\eta })^{\star }\,\boldsymbol{\Psi }-\rho \,\frac{\partial \Omega }{%
\partial \boldsymbol{x}}\,\boldsymbol{\zeta }\right] dv\,dt .
\end{eqnarray*}%
Relations :
\begin{eqnarray*}
(\rho \,\mu -\varepsilon +\eta \,{\mathcal T} )\func{div}\boldsymbol{\zeta } &=&%
\func{div}[(\rho \,\mu -\varepsilon +\eta \,{\mathcal T} )\,\boldsymbol{\zeta }] \\
&+&\left[ \boldsymbol{\Phi }^{\star }\frac{\partial \func{grad}\rho }{%
\partial \boldsymbol{x}}+\boldsymbol{\Psi }^{\star }\frac{\partial \func{grad%
}\eta }{\partial \boldsymbol{x}}-\rho \frac{\partial \mu }{\partial
\boldsymbol{x}}-\eta \frac{\partial {\mathcal T} }{\partial \boldsymbol{x}}\right]
,
\end{eqnarray*}
and
\begin{equation*}
\delta (\func{grad}\rho )^{\star }=\frac{\partial \delta \rho
}{\partial
\boldsymbol{x}}-\frac{\partial \rho }{\partial \boldsymbol{x}}\,\frac{%
\partial \boldsymbol{\zeta }}{\partial \boldsymbol{x}}\quad \mathrm{and}%
\quad \delta (\func{grad}\eta )^{\star }=\frac{\partial \delta \eta }{%
\partial \boldsymbol{x}}-\frac{\partial \eta }{\partial \boldsymbol{x}}\,%
\frac{\partial \boldsymbol{\zeta }}{\partial \boldsymbol{x}},
\end{equation*}
imply
\begin{eqnarray*}
-\,\delta (\func{grad}\rho )^{\star }\,\boldsymbol{\Phi } &=&\func{div}\left[
\boldsymbol{\Phi }\,\frac{\partial \rho }{\partial \boldsymbol{x}}\,%
\boldsymbol{\zeta }-\boldsymbol{\Phi }\,\delta \rho +(\rho \func{div}\,%
\boldsymbol{\Phi })\boldsymbol{\zeta }\right] \\
&&+\left[ \frac{\partial (\rho \func{div}\boldsymbol{\Phi })}{\partial
\boldsymbol{x}}-\func{div}\left( \boldsymbol{\Phi }\,\frac{\partial \rho }{%
\partial \boldsymbol{x}}\right) \right]\boldsymbol{\zeta}\ .
\end{eqnarray*}
and
\begin{eqnarray*}
-\,\delta (\func{grad}\eta )^{\star }\,\boldsymbol{\Psi } &=&\func{div}\left[
\boldsymbol{\Psi }\,\frac{\partial \eta }{\partial \boldsymbol{x}}\,%
\boldsymbol{\zeta }-\boldsymbol{\Psi }\,\delta \eta +(\eta \func{div}\,%
\boldsymbol{\Psi })\boldsymbol{\zeta }\right] \\
&&+\left[ \frac{\partial (\eta \func{div}\boldsymbol{\Psi })}{\partial
\boldsymbol{x}}-\func{div}\left( \boldsymbol{\Psi }\,\frac{\partial \eta }{%
\partial \boldsymbol{x}}\right) \right]\boldsymbol{\zeta}\ .
\end{eqnarray*}%
Consequently,
\begin{equation*}
\rho \,\boldsymbol{u}^{\star }\frac{d\boldsymbol{\zeta }}{dt}=\frac{\partial
}{\partial t}(\rho \,\boldsymbol{u}^{\star }\boldsymbol{\zeta })+\func{div}%
[\rho (\boldsymbol{u}^{\star }\boldsymbol{\zeta })\boldsymbol{u}]-\rho \,%
\boldsymbol{a}^{\star }\boldsymbol{\zeta } ,
\end{equation*}%
where $\boldsymbol{a}$ denotes the acceleration vector of the fluid. We get,
\begin{equation*}
\begin{array}{l}
\delta S=\displaystyle\int_{t_{1}}^{t_{2}}\int_{D_{t}}\left\{ -\rho \,%
\boldsymbol{a}^{\star }-\rho \,\frac{\partial \mu }{\partial \boldsymbol{x}}%
-\eta \,\frac{\partial {\mathcal T} }{\partial \boldsymbol{x}}+\boldsymbol{\Phi }%
^{\star }\,\frac{\partial \func{grad}\rho }{\partial \boldsymbol{x}}+%
\boldsymbol{\Psi }^{\star }\,\frac{\partial \func{grad}\eta }{\partial
\boldsymbol{x}}\right. \label{diffenergy copy(1)} \\
\qquad \qquad \displaystyle\left. +\,\frac{\partial }{\partial \boldsymbol{x}%
}\big( \rho \func{div}\boldsymbol{\Phi }+\eta \func{div}\boldsymbol{\Psi }%
\big) -\func{div}\left( \boldsymbol{\Phi }\,\frac{\partial \rho }{\partial
\boldsymbol{x}}+\boldsymbol{\Psi }\,\frac{\partial \eta }{\partial
\boldsymbol{x}}\right) -\rho \,\frac{\partial \Omega }{\partial \boldsymbol{x%
}}\right\} \,\boldsymbol{\zeta }\,dv\,dt\, \\
\displaystyle+\int_{t_{1}}^{t_{2}}\int_{D_{t}}\left\{ \func{div}\left[
\boldsymbol{\Phi }\,\frac{\partial \rho }{\partial \boldsymbol{x}}\,%
\boldsymbol{\zeta }-\boldsymbol{\Phi }\,\delta \rho +(\rho \func{div}\,%
\boldsymbol{\Phi })\boldsymbol{\zeta +}\boldsymbol{\Psi }\,\frac{\partial
\eta }{\partial \boldsymbol{x}}\,\boldsymbol{\zeta -}\boldsymbol{\Psi }%
\,\delta \eta +(\eta \func{div}\,\boldsymbol{\Psi })\,\boldsymbol{\zeta }%
\right. \right. \\
\left. \qquad \qquad \displaystyle+(\rho \,\mu -\varepsilon +\eta \,{\mathcal T}
)\,\boldsymbol{\zeta }\Big]\,+\frac{\partial }{\partial t}(\rho \,%
\boldsymbol{u}^{\star }\boldsymbol{\zeta })+\func{div}[\,\rho (\boldsymbol{u}%
^{\star }\boldsymbol{\zeta })\boldsymbol{u}\,]\right\} dv\,dt .
\end{array}%
\end{equation*}%
By integration,
the second integral vanishes when the virtual displacement   is null on the boundary of $[t_{1},t_{2}]\times D_{t}$.

 From Hamilton principle,
\begin{equation*}
  \forall  \ \boldsymbol{x}\in D_t\rightarrow \boldsymbol{\zeta } (\boldsymbol{x}), \rm{with}\ \boldsymbol{\zeta } (\boldsymbol{x}) \  \rm{null\ on\ the\ boundary\ of}\  D_{t},\   \delta S =0 ,%
\end{equation*}
we can deduce the motion equation of conservative and isentropic fluids.

\subsection{\label{subsec2.3}First form of motion equation}
From  Hamilton principle, we deduce
\begin{eqnarray}
&&\rho \,\boldsymbol{a}+\rho \,\func{grad}\mu +\eta \,\func{grad}{\mathcal T} \ -%
\frac{\partial \func{grad}\rho }{\partial \boldsymbol{x}}\,\boldsymbol{\Phi }%
\,-\frac{\partial \func{grad}\eta }{\partial \boldsymbol{x}}\,\boldsymbol{%
\Psi }\,  \label{equmotion} \\
&&-\func{grad}\left( \rho \func{div}\boldsymbol{\Phi }+\eta \func{div}%
\boldsymbol{\Psi }\right) +\func{div}{^\star}\left( \boldsymbol{\Phi }\ {\func{grad}}%
^{\star }\rho +\boldsymbol{\Psi }\ {\func{grad}}^{\star }\eta
\right) +\rho\, \func{grad}\Omega =0 . \notag
\end{eqnarray}%
From   relations
\begin{eqnarray*}
&&\frac{\partial (\rho \func{div}\boldsymbol{\Phi })}{\partial \boldsymbol{x}}=%
(\func{div}\boldsymbol{\Phi })\,\frac{\partial \rho }{\partial \boldsymbol{x}}%
+\rho \,\frac{\partial \func{div}\boldsymbol{\Phi }}{\partial \boldsymbol{x}}%
\quad \mathrm{and}\quad \\
&&\frac{\partial (\eta \func{div}\boldsymbol{\Psi })}{\partial
\boldsymbol{x}}
=(\func{div}\boldsymbol{\Psi })\,\frac{\partial \eta }{\partial \boldsymbol{x}}%
+\eta \,\frac{\partial \func{div}\boldsymbol{\Psi }}{\partial \boldsymbol{x}} ,
\\
&&\func{div}\left( \boldsymbol{\Phi }\,\frac{\partial \rho }{\partial
\boldsymbol{x}}\right) =(\func{div}\boldsymbol{\Phi })\,\frac{\partial \rho }{%
\partial \boldsymbol{x}}+\boldsymbol{\Phi }^{\star }\frac{\partial \func{grad%
}\rho }{\partial \boldsymbol{x}}\quad \mathrm{and} \\
&& \func{div}\left( \boldsymbol{\Psi }\,\frac{\partial \eta }{\partial
\boldsymbol{x}}\right) = (\func{div}\boldsymbol{\Psi })\,\frac{\partial \eta }{%
\partial \boldsymbol{x}}+\boldsymbol{\Psi }^{\star }\frac{\partial \func{grad%
}\eta }{\partial \boldsymbol{x}},
\end{eqnarray*}%
we obtain,
\begin{equation}
\rho \,\boldsymbol{a}+\rho \,\func{grad}(\,\mu-\func{div}\boldsymbol{\Phi}+\Omega) +\eta \,\func{grad}(\,{\mathcal T}-\func{div}%
\boldsymbol{\Psi})  =0 , \label{equmotionbis}
\end{equation}%
or,
\begin{equation}
\boldsymbol{a}+\ \func{grad}(\mu-\func{div}\boldsymbol{\Phi}+\Omega)+ s\, \func{grad}({\mathcal T}-\func{div}\boldsymbol{\Psi}) =0 ,\label{motion5}
\end{equation}
where  $s = \eta/\rho$.\\
From Eq. (\ref{equmotionbis}) we deduce,
\begin{equation}
\boldsymbol{a}+\ \func{grad} \Xi -\theta \,\func{grad} s =0  \label{motion4}
\end{equation}
with
\begin{equation*}
 \theta = {\mathcal T} -\func{div}\boldsymbol{\Psi}\qquad \mathrm{and}\qquad \Xi = \mu-\func{div}\boldsymbol{\Phi} + \Omega
+({\mathcal T} -\func{div}\boldsymbol{\Psi})\, s .
\end{equation*}
\emph{In   case of internal energy} (\ref{specialvolumeenergy}) we get
\begin{equation*}
\theta= {\mathcal T}   - D\,\Delta\rho - E\,\Delta\eta \qquad \mathrm{and}\qquad \Xi = \mu- C\,\Delta\rho-D\,\Delta\eta+\Omega
+({\mathcal T} - D\,\Delta\rho-E\,\Delta\eta)\, s ,
\end{equation*}
where $\Delta$ is the Laplace operator.
From $\displaystyle d\mu
=\frac{d{\mathcal P}}{\rho}  -s\,d{\mathcal T} $,  Eq. (\ref{equmotionbis}) can be written
\begin{equation}
\rho \,\boldsymbol{a}+{\func{grad}}\,{\mathcal P}+\rho \,\func{grad}\left(\, \Omega -\func{%
div}\boldsymbol{\Phi }\,\right) -\eta\, \func{grad}\func{div}\boldsymbol{\Psi }=0 .\label{equ acc}
\end{equation}

\subsection{\label{subsec2.4}Second form of  the motion equation}

If we denote
$$
p=\mathcal{P}-\rho \func{div}\boldsymbol{\Phi} -\eta \func{div%
}\boldsymbol{\Psi} \quad {\rm and}\quad
\boldsymbol{\sigma }=-p\, \boldsymbol{I} -\boldsymbol{\Phi} \,\frac{\partial
\rho }{\partial \boldsymbol{x}}-\boldsymbol{\Psi} \,\frac{\partial \eta }{%
\partial \boldsymbol{x}} ,
$$
equation (\ref{equmotion}) yields
\begin{equation}
\rho \,\boldsymbol{a}\ -\ {\func{div}}^{\star }\boldsymbol{\sigma }+\rho\,
\func{grad}\Omega =0 . \label{equmation1}
\end{equation}

In the case of  internal energy (\ref{specialvolumeenergy}), we obtain the
value of $\boldsymbol{\sigma}$,
\begin{equation}
\boldsymbol{\sigma }=-p\ {\boldsymbol I}-(C\func{grad}\rho +D\func{grad}\eta )\,\frac{%
\partial \rho }{\partial \boldsymbol{x}}-(D\func{grad}\rho +E\func{grad}\eta
)\,\frac{\partial \eta}{\partial \boldsymbol{x}} . \label{stresstensor}
\end{equation}

\subsection{Adiabatic motions}

If the total entropy of the fluid in domain $D_{t}$
is constant, its variation  is null,
\begin{equation*}
\delta \int_{D_{t}}\eta \,dv\equiv \delta \int_{D_{t}}\rho s\,dv=0 ,
\end{equation*}%
and it exists a constant Lagrange multiplier $T_{0}$ such that the variation of Hamilton's action
\begin{equation*}
\delta S\equiv\int_{t_{1}}^{t_{2}}\int_{D_{t}}\rho \,\delta \left( \frac{1}{2}\,%
\boldsymbol{u}^{\star }\boldsymbol{u}-\frac{\varepsilon }{\rho }-\Omega
+T_{0}\,s\right) dv\,dt=0   \label{Action2}
\end{equation*}
 is null,
with always
\begin{equation*}
\delta \boldsymbol{u}=\frac{d\boldsymbol{\zeta }}{dt}\quad \rm{and} \quad \delta \rho
=-\rho \,\func{div}\boldsymbol{\zeta } .  \label{variations2}
\end{equation*}%
From variation field $\boldsymbol{\zeta }$, we get the same equation of motions
 (Eq. (\ref{equmotionbis})).\newline When $\boldsymbol{\zeta }=0$,
 independent  variation of $\eta$ \ ($\delta \eta =\rho
\,\delta s$)\   yields
\begin{equation*}
\delta S=\int_{t_{1}}^{t_{2}}\int_{D_{t}}\left( -\,\delta \varepsilon
+\rho\,  T_{0}\,\delta s \right) \ dv\,dt=0.
\end{equation*}
Due to Eq. (\ref{diffenergy}), $\delta \varepsilon ={\mathcal T}\, \delta \eta -\boldsymbol{\Psi}^\star\,\delta \func{grad}\, \eta$\ \ and $\boldsymbol{\zeta
}=0$ \,implies $\displaystyle\delta \func{grad}\eta =\left(\frac{\partial \delta \eta
}{\partial \boldsymbol{x}}\right)^\star$. Consequently,
\begin{eqnarray*}
\delta S &=&\int_{t_{1}}^{t_{2}}\int_{D_{t}}\left[ (T_0 - {\mathcal T})\, \delta \eta -\boldsymbol{\Psi}^\star\,\delta \func{grad}\, \eta\right] \ dv\,dt    \\
&=&\int_{t_{1}}^{t_{2}}\int_{D_{t}}\left( T_{0} - {\mathcal T}+\func{div}\boldsymbol{\Psi}    \right)\ \delta \eta\
dv\,dt-\int_{t_{1}}^{t_{2}}\int_{D_{t}} \func{div}\left( \boldsymbol{\Psi}\,
\delta \eta \right) dv\,dt .
\end{eqnarray*}
 We consider that  $\delta \eta=0$ on the boundary of $D_t$. By integration on
the $D_t$-boundary, the second integral is null and the Hamilton principle yields :
\begin{equation*}
{\mathcal T} = T_{0}+  \func{div}\boldsymbol{\Psi} \,, \label{equtemp0}
\end{equation*}
and in   the special case of   a volume energy in form   (\ref%
{specialvolumeenergy}),
\begin{equation}
 {\mathcal T}  = T_{0}+    D\,\Delta\rho + E\,\Delta\eta \,. \label{equtemp}
\end{equation}
We note that $\theta={\mathcal T}-  \func{div}\boldsymbol{\Psi} $ is constant equal to $T_0$ which is the temperature in the homogeneous parts of   thermocapillary fluids (corresponding to the bulks).

\section{Equation of energy and   second law of thermodynamics \cite{Muller,ET}}

\subsection{Equation of motions of viscous thermocapillary fluids}
For a viscous fluid, we add a  stress tensor in the Newtonian
form
\begin{equation*}
\boldsymbol{\sigma }_{v} = \tau\, ({\rm tr}\, \mathcal{D})\, \boldsymbol{%
I}+  2\,\kappa\, \mathcal{D} ,
\end{equation*}
where $\mathcal{D}$ is the velocity deformation tensor; $\tau, \kappa$ are constant. We are in first gradient model for the viscosity but experiments prove that such a model is always correct for capillary layers \cite{Bocquet}. The Hamilton principle becomes the principle of virtual powers (or virtual works) \cite{Gouin4}
 and Eq. (\ref{equmation1}) allows  to obtain
\begin{equation*}
\rho \,\boldsymbol{a}\ -  {\func{div}}^{\star }\left(\boldsymbol{\sigma }+\boldsymbol{\sigma }_{v}\right)+\rho
\func{grad}\Omega =0  \label{equmation2} ,
\end{equation*}
where $\boldsymbol{\sigma }$ verifies Eq. (\ref{stresstensor}).
\subsection{Equation of energy}
We extend the results proposed in \cite{Eglit,Dunn,Gouin5,trusk}.
Let us note
\begin{equation}
\left\{
\begin{array}{l}
\displaystyle\quad \boldsymbol{M} = \rho \,\boldsymbol{a}-{\func{div}}%
^{\star }\left( \boldsymbol{\sigma }+\boldsymbol{\sigma }_{v}\right) +\rho
\func{grad}\Omega \\
\displaystyle\quad B = \frac{\partial \rho }{\partial t}+\func{div}\rho \,%
\boldsymbol{u} \label{system1} \\
\displaystyle\quad N = \rho \left( {\mathcal T} -\func{div}\boldsymbol{\Psi}
\right) \dot{s}+\func{div}\boldsymbol{q}-r-tr\left(\boldsymbol{\sigma }%
_{v}\mathcal{D}\right) \\
\displaystyle\quad F = \frac{\partial e}{\partial t}+\func{div}\left[ \left(
e \boldsymbol{I}-\boldsymbol{\sigma }-\boldsymbol{\sigma }_{v}\right) \boldsymbol{u}\right]
-\func{div}\left( \dot{\rho}\,\boldsymbol{\Phi} +\dot{\eta}\,\boldsymbol{\Psi%
} \right) +\func{div}\boldsymbol{q}-r-\rho \frac{\partial \Omega }{\partial t%
}%
\end{array}%
\right.
\end{equation}
where $\displaystyle e=\frac{1}{2}\,\rho \,\boldsymbol{u}^{\star }%
\boldsymbol{u}+\varepsilon +\rho \,\Omega $ is the total volume energy of
the fluid, $\boldsymbol{q}$ and $r$ are the heat flux vector and the heat supply, respectively; superscript\ \ $\boldsymbol{\dot{}}$\ \ denotes the material derivative and  the free enthalpy is $\displaystyle h\equiv\frac{\varepsilon +p}{\rho }$. We get :

{\theorem
For an internal energy in form (\ref{volumeenergy}) and for any motion  of
thermocapillary fluids,
\begin{equation}
F-\boldsymbol{M}^{\star }\boldsymbol{u}-\left( \frac{1}{2}\,\boldsymbol{u}%
^{\star }\boldsymbol{u}+h+\Omega \right) B-N\equiv 0  \label{identity1} .
\end{equation}%
}
The proof is proposed in Appendix 1.
\newline
{\corollary For any motion of conservative thermocapillary fluids, the
conservation of specific entropy $\dot{s}=0$ (or ${\partial \eta }/{\partial
t}+\func{div}\eta\, \boldsymbol{u}=0$) is equivalent to
\begin{equation*}
\frac{\partial e}{\partial t}+\func{div}\left[ \left( e \boldsymbol{I}-\boldsymbol{\sigma }%
\right) \boldsymbol{u}\right] -\func{div}\left( \dot{\rho}\,\boldsymbol{\Phi}
+\dot{\eta}\,\boldsymbol{\Psi} \right) -\rho \,\frac{\partial \Omega }{%
\partial t}=0 .
\end{equation*}
}
{\corollary For any motions of dissipative thermocapillary fluids,
equation of energy
\begin{equation*}
\frac{\partial e}{\partial t}+\func{div}\left[ \left( e \boldsymbol{I}-\boldsymbol{\sigma }-%
\boldsymbol{\sigma }_{v}\right) \boldsymbol{u}\right] -\func{div}\left( \dot{%
\rho}\,\boldsymbol{\Phi} +\dot{\eta}\,\boldsymbol{\Psi} \right) +\func{div}%
\boldsymbol{q}-r-\rho \frac{\partial \Omega }{\partial t}=0
\end{equation*}%
is equivalent to   "equation of entropy"
\begin{equation}
\rho \left( {\mathcal T} -\func{div}\boldsymbol{\Psi} \right) \dot{s}+\func{div}%
\boldsymbol{q}-r-tr\left( \boldsymbol{\sigma }_{v}\mathcal{D}\right)=0 .
\label{eqentropy}
\end{equation}
}
Term $\dot{\rho}\,\boldsymbol{\Phi} +\dot{\eta}\,\boldsymbol{\Psi} $ has
the physical dimension of a heat flux vector; it corresponds to the
\emph{interstitial working term} \cite{Dunn} and reveals the
existence of an additional term to the heat flux even if the motion is
conservative. The result extends the ones obtained for capillary fluids when
terms associated with $\func{grad}\eta$ are not taken into account.

\subsection{Planck and Clausius-Duhem inequalities}

For any motion  of thermocapillary fluids, $tr\left( \boldsymbol{\sigma }%
_{v}\mathcal{D}\right)\geq 0 $ \cite{Muller}. Equation (\ref{eqentropy}) implies \emph{ Planck's
inequality} \cite{Truesdell}
\begin{equation*}
\rho \left( {\mathcal T} -\func{div}\boldsymbol{\Psi} \right) \dot{s}+\func{div}%
\boldsymbol{q}-r\geq 0 .
\end{equation*}
We assume the
\emph{Fourier law} in the general form,
\begin{equation*}
\boldsymbol{q}^\star \func{grad} \theta \leq 0,\qquad {\rm with} \qquad   \theta = {\mathcal T} -\func{div}\boldsymbol{\Psi}
\end{equation*}
and we obtain
\begin{equation*}
\rho\, \dot{s}+\func{div}\left(\frac{\boldsymbol{q}}{\theta}\right)-\frac{r}{\theta}\geq 0 ,
\label{CD}
\end{equation*}
 which is the extended form  for
thermocapillary fluids of   \emph{Clausius-Duhem's inequality}. We note that temperature $\theta$ corresponds to  the temperature value in   homogeneous parts of   thermocapillary fluids.

\section{Some properties of thermocapillary fluids}

\subsection{First integrals and Kelvin's circulation-theorems \cite{Landau}}

{\theorem
The velocity circulation on a closed, isentropic fluid-curve is constant.}

The circulation of velocity vector $\boldsymbol{u}$ on a closed fluid-curve $\mathcal{C}$ is $\displaystyle\mathcal{%
J }=\oint_{\mathcal{C}} \boldsymbol{u}^\star\, d\boldsymbol{x}$. From    \cite{Serrin}  p. 162,
\begin{equation*}
\frac{d}{dt}\oint_{\mathcal{C}} \boldsymbol{u}^\star\, d\boldsymbol{x}=\oint_{%
\mathcal{C}} \boldsymbol{a}^\star\, d\boldsymbol{x}
\end{equation*}
and thanks to Eq. (\ref%
{motion4}), we deduce
\begin{equation*}
\oint_{%
\mathcal{C}} \boldsymbol{a}^\star\, d\boldsymbol{x} =
\oint_{\mathcal{C}}{\func{grad}}^\star  {\Xi}\  d\boldsymbol{x} = 0,
\end{equation*}
which proves the theorem. {\corollary In a homentropic motion (the
entropy is uniform in the fluid), the velocity circulation  on a
fluid-curve is constant.} {\theorem
The velocity circulation on a closed  fluid-curve such that ${\mathcal T} -\func{%
div}\boldsymbol{\Psi}= T_0$ is constant.}

From Eq. (\ref{motion5}) we get,
\begin{equation*}
\boldsymbol{a}   + \func{grad}  {\Xi} -( {\mathcal T} -\func{div}%
\boldsymbol{\Psi} ) \func{grad} s = 0 \,.
\end{equation*}
But,
\begin{equation*}
\boldsymbol{a}  -\frac{1}{2}\func{grad}   \boldsymbol{u}^2   = \frac{\partial\boldsymbol{u}}{\partial t}+ \frac{\partial%
\boldsymbol{u}}{\partial \boldsymbol{x}}\, \boldsymbol{u} -\left(\frac{%
\partial\boldsymbol{u}}{\partial \boldsymbol{x}}\right)^\star\boldsymbol{u}=
\frac{\partial\boldsymbol{u}}{\partial t}+ \func{rot} \boldsymbol{u} \times
\boldsymbol{u}\, .
\end{equation*}
For a stationary motion,
\begin{equation}
\func{rot} \boldsymbol{u} \times \boldsymbol{u} = ( {\mathcal T} -\func{div}%
\boldsymbol{\Psi} ) \func{grad} s - \func{grad}\left( \Xi
+\frac{\boldsymbol{u}^2}{2}+\Omega \right).  \label{Crocco}
\end{equation}
Equation(\ref{Crocco}) is the generalized Crocco-Vazsonyi relation for
thermocapillary fluids.

\subsection{Superficial tension of thermocapillary fluids}

We consider a planar interface between liquid and vapour bulks of a
thermocapillary fluid. In the interfacial layer, density gradients are
important. With    internal energy   (\ref{specialvolumeenergy}),
the stress tensor   is
\begin{eqnarray*}
\boldsymbol{\sigma } &=&-\left( \mathcal{P}- \frac{C}{2}\,{\func{grad}}^{\star }\rho \,\func{%
grad}\rho -D\,{\func{grad}}^{\star }\rho \,\func{grad}\eta -\frac{E}{2}\,{%
\func{grad}}^{\star }\eta \,\func{grad}\eta \right)\, \boldsymbol{%
I} \\
&-&(C\func{grad}\rho +D\func{grad}\eta )\,\frac{\partial \rho }{\partial
\boldsymbol{x}}-(D\func{grad}\rho +E\func{grad}\eta )\,\frac{\partial \eta }{%
\partial \boldsymbol{x}}\, .
\end{eqnarray*}
When the extraneous force potential is neglected, the equation of the equilibrium is
\begin{equation*}
\func{div}\boldsymbol{\sigma } =0 .
\end{equation*}%
For a flat interface, normal to\, \ $\func{grad}\rho $\, \ and\, \ $\func{grad}\eta $,    the coordinate normal to the interface being denoted $z$,
the eigenvalues of stress tensor $\boldsymbol{\sigma } $ are%
\begin{equation*}
\lambda _{1}=-\mathcal{P}+\frac{C}{2}\,
\left(\frac{d\rho}{dz}\right)^2 +D\,\frac{d\rho}{dz}\,\frac{d\eta}{dz} +\frac{E}{2}\,
\left(\frac{d\eta}{dz}\right)^2
\end{equation*}%
(associated with the plane of interface), and%
\begin{equation*}
\lambda _{2}=-\mathcal{P}- \frac{C}{2}\,
\left(\frac{d\rho}{dz}\right)^2 -D\,\frac{d\rho}{dz}\,\frac{d\eta}{dz} -\frac{E}{2}\,
\left(\frac{d\eta}{dz}\right)^2
\end{equation*}%
(associated with direction normal to the plane of interface). \newline
In an orthonormal system with  third coordinate $z$,
the stress tensor  writes
\begin{equation*}
\boldsymbol{\sigma } =%
\begin{bmatrix}
\lambda _{1} & 0 & 0 \\
0 & \lambda _{1} & 0 \\
0 & 0 & \lambda _{2}%
\end{bmatrix}%
.
\end{equation*}%
The equation of balance momentum in the planar interface implies
\begin{equation*}
\lambda _{2}= -  {P}_{0}\,,
\end{equation*}%
where ${P}_{0}$ is the common pressure in the bulks. The force per unit of length on the edge of the interface is (see Fig.  1) :
\begin{equation*}
\mathcal{F}=\int_{z_{1}}^{z_{2}}\lambda _{1}\,dz=
\end{equation*}%
\begin{equation*}
- P_{0}(z_{2}-z_{1})+\int_{z_{1}}^{z_{2}}\left[ C\,
\left(\frac{d\rho}{dz}\right)^2 +2\,D\ \frac{d\rho}{dz}\,\frac{d\eta}{dz} +E\,
\left(\frac{d\eta}{dz}\right)^2 \right]
\ dz,
\end{equation*}%
where $z_{2}-z_{1}$ corresponds to the  physical interface
thickness. Due to the small thickness of the interface,
$-P_{0}(z_{2}-z_{1})$ is negligible. Let us note
\begin{equation*}
 H_{1} = \int_{z_{1}}^{z_{2}}C\,\left(\frac{d\rho}{dz}\right)^2 \ dz, \quad
H_{2} = \int_{z_{1}}^{z_{2}}2\,D\, \frac{d\rho}{dz} \,\frac{d\eta}{dz}\ dz,  \quad  H_{3}=\int_{z_{1}}^{z_{2}}E\,
\left(\frac{d\eta}{dz}\right)^2  \ dz \ .
\end{equation*}%
The line force per unit of length on the interface edge is
\begin{equation*}
 H=H_{1}+H_{2}+H_{3} ,
\end{equation*}%
where $H$ represents the surface tension of the planar interface at equilibrium.%
\begin{figure}[h]
\begin{center}
\includegraphics[width=10cm]{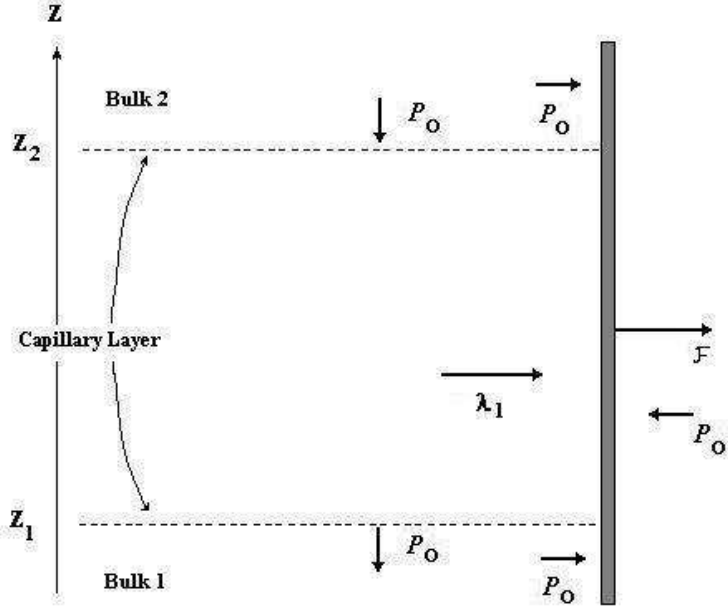}
\end{center}
\caption{Interpretation of the surface tension}
\label{fig1}
\end{figure}
If we consider the approximation
\begin{equation}
\frac{\partial \varepsilon _{_{0}}}{\partial \eta }(\rho ,\eta )=T
_{1},  \label{rhoeta}
\end{equation}
where $T_1$ is the  temperature value in the liquid and vapour bulks, then $\eta $ is a
function of $\rho $. Due to the variation principle, the surface tension calculated for capillary fluids (corresponding to $D=0$ and $E=0$) with   approximation (\ref{rhoeta}) is necessary greater than the surface tension when
\begin{equation*}
 \frac{\partial \varepsilon _{_0}}{\partial \eta }%
(\rho ,\eta )=T_{1}+D\,\Delta \,\rho +E\,\Delta \,\eta\, .
\end{equation*}%
 In fact,
experiments prove that the entropy effects are small enough on surface tension value and when the critical point is approached, the one - and two-density theories become equivalent  as a general property of  critical point (\cite{rowlinson}, Ch. 3), \cite{domb}.

\subsection{Maxwell's rule for thermocapillary fluids}
We consider the case when the volume internal energy is
 in  form (\ref{specialvolumeenergy}). In the case of capillary fluids
 (corresponding to $D = 0$ and $E = 0$),  the  Maxwell rule of
 planar liquid-vapour interface at equilibrium can be written in equivalent form
\begin{equation*}
 \int_{\rho_v}^{\rho_l} \left(\mu_{0}-\mu_{1}\right) d\rho = 0 ,
\end{equation*}
where $\rho_l$ and $ \rho_v$ are the   mass density in the liquid and vapour bulks; $\mu_{1}$ is the common value of the chemical potential in the bulks  \cite{Rocard}. We denote $\eta_l$ and $\eta_v$  the volume entropies in the liquid and vapour bulks, respectively.

 Equation of temperature (\ref{equtemp}) of thermocapillary fluids yields
\begin{equation*}
{\mathcal T}- T_{1}= D\,\frac{d^2\rho}{dz^2}+E\,\frac{d^2\eta}{dz^2}\,. \label{temp}
\end{equation*}
Without body forces,    equation of equilibrium (\ref{motion5}) of thermocapillary fluids    yields
\begin{equation*}
\func{grad}(\mu-\func{div}\boldsymbol{\Phi}) = 0 \label{equi}
\end{equation*}
or by integration,
\begin{equation*}
\mu- \mu_{1}= C\,\frac{d^2\rho}{dz^2}+D\,\frac{d^2\eta}{dz^2}\, . \label{equil}
\end{equation*}

Consequently,
\begin{eqnarray*}
&&\int_{\rho_v}^{\rho_l}(\mu-\mu_{1})\, d\rho+\int_{\eta_v}^{\eta_l}({\mathcal T}- T_{1})\, d\eta   \\
&=&\int_{z_{1}}^{z_{2}}\left[ C\, \frac{d^2\rho}{dz^2}\,\frac{d \rho}{dz} + D\left(  \frac{d^2\eta}{dz^2}\,\frac{d \rho}{dz}+ \frac{d^2\rho}{dz^2}\,\frac{d \eta}{dz}\right)+E \frac{d^2\eta}{dz^2}\,\frac{d \eta}{dz}\right] dz  \\
&=& \qquad \left[\frac{C}{2}\left(\frac{d\rho}{dz}\right)^2 + D \left(\frac{d\rho}{dz}\right)\left(\frac{d\eta}{dz}\right)+
\frac{E}{2}\left(\frac{d\eta}{dz}\right)^2\right]_{z_{1}}^{z_{2}} \equiv 0\, .
\end{eqnarray*}
The generalisation of   Maxwell's rule    for thermocapillary fluids writes in the form :
\begin{equation*}
\int_{\rho_v}^{\rho_l}(\mu-\mu_{1})\, d\rho+\int_{\eta_v}^{\eta_l}({\mathcal T}- T_{1})\, d\eta = 0\,.
\end{equation*}

\section{Governing equations in divergence form and hyperbolicity}

Conservative motions with balance equation of energy lead to  an
 interesting class of quasilinear systems previously pointed out by Godunov \cite{Godunov}, Friedrichs and Lax  \cite{Friedrichs}. In classical mechanics and relativity, many studies on hyperbolic systems were developed in the literature for hydrodynamics, elasticity   and classical materials  \cite{Boillat,Boillat2,Ruggeri2,Ruggeri1}. The section  extends   results presented in \cite{Gavrilyuk2} for the capillary-fluids' simplest case. The small motions near an equilibrium position are studied thanks to  a convenient system  of governing equations  associated with a Legendre transformation of the internal energy.
\subsection{Governing equations in divergence form}

Let us denote   $\boldsymbol{\beta }\equiv\func{grad}\,\rho $ ,     $\boldsymbol{%
\chi }\equiv\func{grad}\,\eta $ and $\boldsymbol{j} \equiv \rho \,\boldsymbol{u}$.
The gradient of the mass-conservation balance  verifies another conservation
equation,
\begin{equation}
\frac{\partial \boldsymbol{\beta }}{\partial t}+{\mathop{\rm grad}}%
\mathop{\rm div}\boldsymbol{j}=0 \label{Beta},
\end{equation}%
Conversely, if we consider $\boldsymbol{\beta }$ as an independent vector verifying Eq. (\ref{Beta}), and if we add   initial condition
\begin{equation*}
\boldsymbol{\beta }\,|_{\,t=0}={\mathop{\rm grad}}\,\rho \,|_{\,t=0}\ ,
\end{equation*}%
then  $\boldsymbol{\beta }\equiv\mathrm{{grad}\,\rho }$ becomes a consequence of
  governing equation (\ref{Beta}).\newline
Similarly, the gradient of the balance of entropy verifies another
conservation equation,
\begin{equation}
\frac{\partial \boldsymbol{\chi }}{\partial t}+{\mathop{\rm grad}}%
\mathop{\rm div}(\eta \,\boldsymbol{u})=0 \label{Chi}.
\end{equation}
In the same way, if we add   initial condition
\begin{equation*}
\boldsymbol{\chi }\,|_{\,t=0}={\mathop{\rm grad}}\,\eta \,|_{\,t=0}\ ,
\end{equation*}
  $\boldsymbol{\chi}\equiv\mathrm{{grad}\,\eta}$ becomes a consequence of
  governing equation (\ref{Chi}) and
we can consider $\boldsymbol{\chi }$ as an independent vector verifying Eq. (\ref{Chi}).\newline  Without body forces, with the new notations, Eqs. (\ref{mass conservation}, \ref{entropy conservation}, \ref{equ acc}, \ref{Beta}, \ref{Chi}) immediately yield the system of governing equations
in the form
\begin{equation}
\left\{
\begin{array}{l}
\quad \displaystyle\frac{\partial \rho }{\partial t}+\mathop{\rm div}%
\boldsymbol{j}=0 \\
\quad \displaystyle\frac{\partial \eta }{\partial t}+\mathop{\rm div}\left( %
\displaystyle\frac{\eta }{\rho }\ \boldsymbol{j}\right) =0\label%
{systemnhyperbolic2} \\
\quad \displaystyle\frac{\partial \boldsymbol{j}^{\star }}{\partial t}+%
\mathop{\rm div}\left( \displaystyle\frac{\boldsymbol{jj}^{\star }}{\rho }+%
\mathcal{P}\,\boldsymbol{I}\right) -\rho \ {\mathop{\rm grad}}^{\star }\left( %
\mathop{\rm div}\ \boldsymbol{\Phi }\right) -\eta \,{\func{grad}}^{\star }%
\left(\func{div}\boldsymbol{\Psi }\right)=0 \\
\quad \displaystyle\frac{\partial \boldsymbol{\beta }}{\partial t}+{%
\mathop{\rm grad}}\left(\mathop{\rm div}\boldsymbol{j}\right)=0 \\
\quad \displaystyle\frac{\partial \boldsymbol{\chi }}{\partial t}+{%
\mathop{\rm grad}}\mathop{\rm div}(\frac{\eta }{\rho }\,\boldsymbol{j})=0\,.
\end{array}%
\right.
\end{equation}

With the new notations, the total volume energy
of the fluid is
\begin{equation*}
{\mathcal E}= \frac{\boldsymbol{j}^{\star }\boldsymbol{j}}{2\,\rho }+ \varepsilon \,.
\end{equation*}%
We denote $\displaystyle q=\mu -\frac{\boldsymbol{u}^{\star }\boldsymbol{u}}{2}
$. Consequently,
\begin{equation*}
d{\mathcal E}=q\,d\rho + {\mathcal T} \,d\eta +\boldsymbol{u}^{\star }d%
\boldsymbol{j}+\boldsymbol{\Phi }^{\star }d\boldsymbol{\beta }+%
\boldsymbol{\Psi }^{\star }d\boldsymbol{\chi } .
\end{equation*}%
The Legendre transform of ${\mathcal E}$ with respect to $\rho ,\eta ,%
\boldsymbol{j},\boldsymbol{\beta },\boldsymbol{\chi }$ is
\begin{equation}
\Pi =\rho \,q+\eta \,{\mathcal T} +\boldsymbol{j}^{\star }\boldsymbol{u}+%
\boldsymbol{\Phi }^{\star }\boldsymbol{\beta }+\boldsymbol{\Psi }^{\star }%
\boldsymbol{\chi }-{\mathcal E} .\label{Legendre}
\end{equation}
Conjugate variables $q,{\mathcal T} ,\boldsymbol{u},\boldsymbol{%
\Phi },\boldsymbol{\Psi }$ verify
\begin{equation*}
\frac{\partial \Pi }{\partial q}=\rho ,\quad \frac{\partial \Pi }{\partial {\mathcal T}}%
=\eta ,\quad \frac{\partial \Pi }{\partial \boldsymbol{u}}=\boldsymbol{j}%
^{\star },\quad \frac{\partial \Pi }{\partial \boldsymbol{\Phi }}=%
\boldsymbol{\beta }^{\star },\quad \frac{\partial \Pi }{\partial \boldsymbol{%
\Psi }}=\boldsymbol{\chi }^{\star } .
\end{equation*}%
System (\ref{systemnhyperbolic2}) can be written in   divergence form as (See
Appendix B) :
\begin{equation}
\left\{
\begin{array}{l}
\quad \displaystyle\frac{\partial }{\partial t}\left( \frac{\partial \Pi }{%
\partial q}\right) +\mathop{\rm div}\left[ \frac{\partial (\Pi \boldsymbol{u)%
}}{\partial q}\right] =0 \\
\quad \displaystyle\frac{\partial }{\partial t}\left( \frac{\partial \Pi }{%
\partial {\mathcal T} }\right) +\mathop{\rm div}\left[ \frac{\partial (\Pi
\boldsymbol{u)}}{\partial {\mathcal T} }\right] =0 \\
\displaystyle\quad \frac{\partial }{\partial t}\left( \frac{\partial \Pi }{%
\partial \boldsymbol{u}}\right) +\mathop{\rm div}\left[ \frac{\partial (\Pi
\boldsymbol{u)}}{\partial \boldsymbol{u}}-\frac{\partial \Pi }{\partial q}%
\dfrac{\partial \boldsymbol{\Phi }}{\partial \boldsymbol{x}}-\frac{\partial
\Pi }{\partial {\mathcal T} }\dfrac{\partial \boldsymbol{\Psi }}{\partial
\boldsymbol{x}}\right] =0\label{system4} \\
\displaystyle\quad \frac{\partial }{\partial t}\left( \frac{\partial \Pi }{%
\partial \boldsymbol{\Phi }}\right) +\mathop{\rm div}\left[ \frac{\partial
(\Pi \boldsymbol{u)}}{\partial \boldsymbol{\Phi }}+\frac{\partial \Pi }{%
\partial q}\dfrac{\partial \boldsymbol{u}}{\partial \boldsymbol{x}}\right] =0
\\
\displaystyle\quad \frac{\partial }{\partial t}\left( \frac{\partial \Pi }{%
\partial \boldsymbol{\Psi }}\right) +\mathop{\rm div}\left[ \frac{\partial
(\Pi \boldsymbol{u)}}{\partial \boldsymbol{\Psi }}+\frac{\partial \Pi }{%
\partial {\mathcal T} }\dfrac{\partial \boldsymbol{u}}{\partial \boldsymbol{x}}%
\right] =0\,.%
\end{array}%
\right.
\end{equation}
When $\varepsilon = \varepsilon (\rho, \eta)$, we get the classical gas
dynamics equations and the conservative form of Godunov \cite{Godunov}.  In the simplest special case, when $\displaystyle\varepsilon = \varepsilon_{_0} (\rho, \eta)+\frac{ C}{2} (\func{grad} \rho)^2$, we obtain the results   \cite{Gavrilyuk2}.

\subsection{Hyperbolicity of governing equations}
The system of  governing equations  generates dispersive relations with multiple eigenvalues near an equilibrium position. In this subsection we extend the results presented in \cite{Swift,RMSeccia,BLR}.
System (\ref{system4}) yields constant solutions
$$
(\,\rho _{e},\eta _{e},%
\boldsymbol{j}_{e},\boldsymbol{\beta}_{e}=0, \boldsymbol{\chi}_{e}=0
\,),
$$
where subscript $e$ means at equilibrium. Since the governing
equations are invariant under Galilean transformation, we can
assume that $\boldsymbol{u}_{e}=0$  which implies
$\boldsymbol{j}_{e}=0$.
\newline
Near equilibrium, we look for
 the solutions proportional to $\displaystyle e^{i\left( \boldsymbol{k}^{\star }\boldsymbol{x%
}-\lambda t\right) }$, where $\boldsymbol{k}^\star =[k_1,k_2,k_3]$ is a constant covector, $\lambda$ a constant scalar and $i^2 =-1$,
\begin{equation*}
\boldsymbol{v} = \boldsymbol{v}_{_0}e^{i\left( \boldsymbol{k}^{\star }%
\boldsymbol{x}-\lambda t\right) } \  \mathrm{with}\ \boldsymbol{v}%
^\star = \left[q, {\mathcal T}, \boldsymbol{u}, \boldsymbol{\Phi}, \boldsymbol{%
\Psi}\right]\ \mathrm{and} \ \boldsymbol{v}_{_0}^\star =
\left[q_{_0},
{\mathcal T}_{_0}, \boldsymbol{u}_{_0}, \boldsymbol{\Phi}_{_0}, \boldsymbol{\Psi}%
_{_0}\right]\,. \label{perturbation}
\end{equation*}
We obtain
\begin{equation*}
\frac{\partial}{\partial t}\left( \frac{\partial \Pi }{\partial \boldsymbol{v%
}}\right)_{e}^\star \equiv \frac{\partial}{\partial \boldsymbol{v}}\left( \frac{%
\partial \Pi }{\partial \boldsymbol{v}}\right)_{e}^\star\frac{\partial \boldsymbol{%
v}}{\partial t} \equiv -i\,\lambda \,\frac{\partial}{\partial \boldsymbol{v}}%
\left( \frac{\partial \Pi }{\partial \boldsymbol{v}}\right)_{e}^\star \boldsymbol{v%
}_{_0}\, e^{i\left( \boldsymbol{k}^{\star }\boldsymbol{x}-\lambda
t\right) } .
\end{equation*}

\begin{equation*}
\func{div}\left(\frac{\partial\, \Pi \boldsymbol{u} }{\partial \boldsymbol{v}%
}\right)^\star= \sum_{j=1}^3 \frac{\partial}{\partial {x^j}}\left(\frac{\partial\, \Pi u^j }{\partial \boldsymbol{%
v}}\right)^\star= \sum_{j=1}^3 \frac{\partial}{\partial
\boldsymbol{v}}\left(\frac{\partial\, \Pi u^j }{\partial \boldsymbol{v}}%
\right)^\star \frac{\partial\boldsymbol{v}}{\partial{x^j}},
\end{equation*}
with  $\boldsymbol{x}^\star =[x^1,x^2,x^3]$ and
\begin{equation*}
\func{div}\left(\frac{\partial\, \Pi \boldsymbol{u} }{\partial \boldsymbol{v}%
}\right)_e^\star = \sum_{j=1}^3 i\,{\boldsymbol
F}^j\,k_j\,\boldsymbol{v}_{_0}\,e^{i\left( \boldsymbol{k}^{\star
}\boldsymbol{x}-\lambda t\right) },
\end{equation*}
where
\begin{equation*}
{\boldsymbol F}^j \equiv \frac{\partial}{\partial \boldsymbol{v}}\left(\frac{\partial\,
\Pi u^j }{\partial \boldsymbol{v}}\right)^\star_{e} ; \quad \mathrm{ we \
denote}\quad {\boldsymbol F} \equiv \sum_{j=1}^3 {\boldsymbol F}^j\, k_j .
\end{equation*}
\emph{\textbf{At equilibrium,}}

\quad $\bullet$ For Eq. (\ref{systemnhyperbolic2})$^3$ (or equivalently Eq. (%
\ref{system4})$^3$), we  add two additive terms   to
classical-fluids' equations :

\emph{\ First term,}
\begin{equation*}
\func{div}\left(\frac{\partial \Pi }{\partial q}\dfrac{\partial \boldsymbol{%
\Phi }}{\partial \boldsymbol{x}}\right)=\left({\func{grad}}^\star\,\rho\right)\,\dfrac{%
\partial \boldsymbol{\Phi }}{\partial \boldsymbol{x}}+ \rho\, \func{div}%
\left(\dfrac{\partial \boldsymbol{\Phi }}{\partial \boldsymbol{x}}\right)\,.
\end{equation*}
At equilibrium, $\func{grad}_e \rho =0$
. Then, from $\boldsymbol{\Phi}=\boldsymbol{%
\Phi}_{_0} e^{i\left(\boldsymbol{k}^{\star } \boldsymbol{x}-\lambda
t\right)}$,
\begin{equation*}
\func{div}\left(\frac{\partial \Pi }{\partial q}\dfrac{\partial \boldsymbol{%
\Phi }}{\partial \boldsymbol{x}}\right)_e= \rho_e\, \func{div}\left(\frac{%
\partial \boldsymbol{\Phi}}{\partial\boldsymbol{x}}\right) = i^2\,\rho_e
\boldsymbol{\Phi}_{_0}^\star\boldsymbol{k}\, \boldsymbol{k}^\star e^{i\left(%
\boldsymbol{k}^{\star } \boldsymbol{x}-\lambda t\right)} .
\end{equation*}
\emph{Second term,}
\begin{equation*}
\func{div}\left(\frac{\partial \Pi }{\partial {\mathcal T}}\frac{\partial
\boldsymbol{\Psi }}{\partial \boldsymbol{x}}\right)=\left({\func{grad}}%
^\star\,\eta\right)\,\frac{\partial \boldsymbol{\Psi }}{\partial \boldsymbol{x}}+
\eta\, \func{div}\left(\dfrac{\partial \boldsymbol{\Psi}}{\partial
\boldsymbol{x}}\right)\ .
\end{equation*}
At equilibrium, $\func{grad}_e \eta =0$   %
. Then, from $\boldsymbol{\Psi}=\boldsymbol{%
\Psi}_{_0} e^{i\left(\boldsymbol{k}^{\star } \boldsymbol{x}-\lambda
t\right)}$,
\begin{equation*}
\func{div}\left(\frac{\partial \Pi }{\partial {\mathcal T}}\dfrac{\partial
\boldsymbol{\Psi }}{\partial \boldsymbol{x}}\right)_e= \rho_e\, \func{div}%
\left(\frac{\partial \boldsymbol{\Psi}}{\partial\boldsymbol{x}}\right) =
i^2\,\eta_e \boldsymbol{\Psi}_{_0}^\star\boldsymbol{k}\, \boldsymbol{k}%
^\star e^{i\left(\boldsymbol{k}^{\star } \boldsymbol{x}-\lambda
t\right)} .
\end{equation*}
Taking account of $\ \boldsymbol{u} = \boldsymbol{u}_{_0}\,
e^{i\left( \boldsymbol{k}^{\star } \boldsymbol{x}-\lambda t\right)}
,$

\quad $\bullet$ For Eq. (\ref{system4})$^4$  at equilibrium, we  add
term
\begin{equation*}
\mathop{\rm div}\left[\frac{\partial \Pi }{\partial
q}\dfrac{\partial \boldsymbol{u}}{\partial \boldsymbol{x}}\right]_e
= i^2\,\rho_e\, \boldsymbol{u}_{_0}^{\star }
\boldsymbol{k}\,\boldsymbol{k}^{\star }e^{i\left(
\boldsymbol{k}^{\star } \boldsymbol{x}-\lambda t\right)} .
\end{equation*}

\quad $\bullet $ For Eq. (\ref{system4})$^{5}$   at equilibrium, we add
term
\begin{equation*}
\mathop{\rm div}\left[ \frac{\partial \Pi }{\partial {\mathcal T} }\frac{\partial
\boldsymbol{u}}{\partial \boldsymbol{x}}\right] _{e}=i^{2}\,\eta _{e}\,%
\boldsymbol{u}_{_{0}}^{\star }\boldsymbol{k}\,\boldsymbol{k}^{\star
}e^{i\left( \boldsymbol{k}^{\star }\boldsymbol{x}-\lambda t\right)
}.
\end{equation*}%
Let us define $\boldsymbol{A}, \boldsymbol{C}$ such that
\begin{equation*}
\displaystyle \boldsymbol{A}=\frac{\partial }{\partial \boldsymbol{v}}\left[ \left( \frac{%
\partial \Pi }{\partial \boldsymbol{v}}\right) ^{\star }\right] _{e},
\end{equation*}
\begin{eqnarray*}
\boldsymbol{C} &=&-\boldsymbol{C}^ {\star}=\left[
\begin{array}{ccccc}
0 & \ \ 0 & \boldsymbol{0}^{\star} & \boldsymbol{0}^{\star } & \boldsymbol{0}%
^{\star } \\
0 & \ \ 0 & \boldsymbol{0}^{\star } & \boldsymbol{0}^{\star } & \boldsymbol{0}%
^{\ast } \\
\boldsymbol{0} & \ \ \boldsymbol{0} & \boldsymbol{O} & -\rho _{e}\,\boldsymbol{kk%
}^{\ast } & -\eta _{e}\,\boldsymbol{kk}^{\star } \\
\boldsymbol{0} & \ \ \boldsymbol{0} & \ \ \rho
_{e}\,\boldsymbol{kk}^{\star }
& \boldsymbol{O} & \boldsymbol{O} \\
\boldsymbol{0} & \ \ \boldsymbol{0} & \ \ \eta
_{e}\,\boldsymbol{kk}^{\star }
& \boldsymbol{O}  & \boldsymbol{O}%
\end{array}%
\right] \quad  \\
\end{eqnarray*}
with
\begin{eqnarray*} && \qquad\qquad \boldsymbol{O}  = \left[
\begin{array}{ccc}
0 & 0 & 0 \\
0 & 0 & 0 \\
0 & 0 & 0%
\end{array}%
\right] \quad \text{and}\quad \boldsymbol{0}^{\ast }=[0\ 0\ 0].
\end{eqnarray*}
Due to $\overline{i\,\boldsymbol{C}}^{\;\star }\equiv i\,\boldsymbol{C}$, where overline denotes the complex conjugation; matrix $i\,\boldsymbol{C}$ is hermitian.%
\newline
The solutions corresponding to the perturbations of system (\ref{system4})
 verify :
\begin{equation*}
i\left[ \,\boldsymbol{F}+i\,\boldsymbol{C}-\lambda\,\boldsymbol{A}
\,\right] \boldsymbol{v}_{_{0}}e^{i\left( \boldsymbol{k}^{\star
}\boldsymbol{x}-\lambda t\right) }=0,
\end{equation*}%
where $\boldsymbol{D}=\overline{\boldsymbol{D}}^{\,\star }\equiv \boldsymbol{F}+i\,\boldsymbol{C}$ is   Hermitian matrix and $\boldsymbol{A}$
is   symmetric matrix; so, $\lambda $ are the roots of the
characteristic equation :
\begin{equation*}
\det \left[ \,\boldsymbol{D}-\lambda \,\boldsymbol{A}\,\right] ={0},
\end{equation*}%
and $\lambda $ is eigenvalue of $\boldsymbol{D}$ with respect to $\boldsymbol{A}$ and $%
\boldsymbol{v}_{_{0}}$ is its eigenvector. Near an equilibrium state where   the local  internal energy is locally convex, $\boldsymbol{A}$
is positive definite; eigenvalues  are real   and the small
perturbations are stable with respect to equilibrium positions.

\section{Conclusion}
For conservative processes associated with system (\ref{systemnhyperbolic2}), Legendre transformation (\ref{Legendre}) of the internal energy yields a system of  governing equations which extends \emph{the classical models of hyperbolicity} to   non-local behaviour.
The Lax-Friedrichs method \cite{Press}  is a numerical method   we can consider  as an alternative to Godunov's scheme \cite{Godunov2} in which one avoids solving a Riemann problem at each cell interface, at the expense of adding artificial viscosity.
 The stability of quasi-linear perturbations allows to forecast    an extention of the Lax-Friedrichs method for thermocapillary fluids.

\section{Appendix A: Proof of relation (\protect\ref{identity1})}

 By using System (\ref{system1}) in the first member of Eq. (\ref{identity1}), dissipative terms $\boldsymbol{q},r,\boldsymbol{\sigma }_{v}$
can be  algebraically simplified. Also are terms associated with inertia and $\Omega $.
The remaining terms are
\begin{equation*}
\left\{
\begin{array}{l}
\displaystyle \boldsymbol{M}_{0} =-\,{\func{div}}^{\star }\left( \boldsymbol{%
\sigma }\right) \\
\displaystyle B =\frac{\partial \rho }{\partial t}+\func{div}\rho \,%
\boldsymbol{u} \\
\displaystyle N_{0} =\rho \left( {\mathcal T} -\func{div}\boldsymbol{\Psi }\right)
\dot{s} \\
\displaystyle F_{0} =\frac{\partial \varepsilon}{\partial t}+\func{div}\left[ \left( \varepsilon \boldsymbol{I}-%
\boldsymbol{\sigma }\right) \boldsymbol{u}\right] -\func{div}\left( \dot{\rho%
}\,\boldsymbol{\Phi }+\dot{\eta}\,\boldsymbol{\Psi }\right)\,,
\end{array}%
\right.
\end{equation*}
and we have to prove
\begin{equation}
F_{0}-\boldsymbol{M}_{0}^{\star }\,\boldsymbol{u}-h\,B-N_{0}\equiv 0
\,. \label{algebraic relation}
\end{equation}%
From
\begin{eqnarray*}
&&\frac{\partial \varepsilon }{\partial t}+\func{div}(\varepsilon \,%
\boldsymbol{u})-(\func{div}\boldsymbol{\sigma })\,\boldsymbol{u}-\func{div}%
\left( \dot{\rho}\,\boldsymbol{\Phi }+\dot{\eta}\,\boldsymbol{\Psi }\right) =
\\
&&\frac{\varepsilon +p}{\rho }\,B+{\mathcal P}\,\dot{\rho}\,+\rho \,{\mathcal T} \,\dot{s}+%
\boldsymbol{\Phi }^{\star }\frac{d\func{grad}\rho }{d\boldsymbol{t}}+%
\boldsymbol{\Psi }^{\star }\frac{d\func{grad}\eta }{dt}-p\,\frac{\dot{\rho}}{%
\rho } \\
&&+\boldsymbol{\Phi }^{\star }\left( \frac{\partial \boldsymbol{u}}{\partial
\boldsymbol{x}}\right) ^{\star }\func{grad}\rho +\boldsymbol{\Psi }^{\star
}\left( \frac{\partial \boldsymbol{u}}{\partial \boldsymbol{x}}\right)
^{\star }\func{grad}\eta -\func{div}\left( \dot{\rho}\,\boldsymbol{\Phi }+%
\dot{\eta}\,\boldsymbol{\Psi }\right) = \\
&&\frac{\varepsilon +p}{\rho }\,B+\rho \left( {\mathcal T} -\func{div}\boldsymbol{%
\Psi }\right) \dot{s}+\boldsymbol{\Phi }^{\star }\func{grad}\frac{\partial
\rho }{\partial t}+\boldsymbol{\Phi }^{\star }\,\frac{\partial \func{grad}%
\rho }{\partial \boldsymbol{x}}\boldsymbol{u+}\boldsymbol{\Psi }^{\star }%
\func{grad}\frac{\partial \eta }{\partial t} \\
&&+\boldsymbol{\Psi }^{\star }\frac{\partial \func{grad}\eta }{\partial
\boldsymbol{x}}\boldsymbol{u+\ }\boldsymbol{\Phi }^{\star }\left( \frac{%
\partial \boldsymbol{u}}{\partial \boldsymbol{x}}\right) ^{\star }\func{grad}%
\rho +\boldsymbol{\Psi }^{\star }\left( \frac{\partial \boldsymbol{u}}{%
\partial \boldsymbol{x}}\right) ^{\star }\func{grad}\eta \\
&&-\left({\func{grad}}^{\star }\,\frac{\partial \rho }{\partial t}\right)\,\boldsymbol{%
\Phi }^{\ }-\boldsymbol{u}^{\star }\,\frac{\partial \func{grad}\rho }{%
\partial \boldsymbol{x}}\,\boldsymbol{\Phi }-({\func{grad}}^{\star }\,\rho) \,%
\frac{\partial \boldsymbol{u}}{\partial \boldsymbol{x}}\,\boldsymbol{\Phi }
\\
&&-\left({\func{grad}}^{\star }\,\frac{\partial \eta }{\partial t}\right)\,\boldsymbol{%
\Psi }^{\ }-\boldsymbol{u}^{\star }\,\frac{\partial \func{grad}\eta }{%
\partial \boldsymbol{x}}\,\boldsymbol{\Psi }-({\func{grad}}^{\star }\,\eta) \,%
\frac{\partial \boldsymbol{u}}{\partial \boldsymbol{x}}\,\boldsymbol{\Psi } ,
\end{eqnarray*}%
and
\begin{equation*}
\frac{\partial \func{grad}\rho }{\partial \boldsymbol{x}}=\left( \frac{%
\partial \func{grad}\rho }{\partial \boldsymbol{x}}\right) ^{\star }\quad
\mathrm{and}\quad \frac{\partial \func{grad}\eta }{\partial \boldsymbol{x}}%
=\left( \frac{\partial \func{grad}\eta }{\partial \boldsymbol{x}}\right)
^{\star }
\end{equation*}%
we get,
\begin{equation*}
\frac{\partial \varepsilon }{\partial t}+\func{div}(\varepsilon \,%
\boldsymbol{u})-(\func{div}\boldsymbol{\sigma })\,\boldsymbol{u}-\func{div}%
\left( \dot{\rho}\,\boldsymbol{\Phi }+\dot{\eta}\,\boldsymbol{\Psi }\right) =%
\frac{\varepsilon +p}{\rho }\, B+\rho \left({\mathcal T} -\func{div}\boldsymbol{\Psi }%
\right) \dot{s} .
\end{equation*}%
Relation
\begin{equation*}
\func{div}(\boldsymbol{\sigma }\,\boldsymbol{u})=(\func{div}\boldsymbol{%
\sigma })\,\boldsymbol{u}+tr\left( \boldsymbol{\sigma }\,\frac{\partial
\boldsymbol{u}}{\partial \boldsymbol{x}}\right)
\end{equation*}%
yields relation (\ref{algebraic relation}). \qquad\qquad\qquad\qquad\qquad\qquad\qquad\qquad\qquad\qquad\qquad\qquad$\square$

\section{Appendix B: Proof of relation (\protect\ref{system4})}

$\displaystyle \mathrm{Relations} \quad \frac{\partial \Pi }{\partial q} =
\rho\quad \mathrm{and}\quad \frac{\partial (\Pi \boldsymbol{u})}{\partial q}
=\rho\,\boldsymbol{u}\quad $ imply Eq. (\ref{system4}$^1$).

$\displaystyle \mathrm{Relations} \quad \frac{\partial \Pi }{\partial {\mathcal T}}
= \eta\quad \mathrm{and}\quad \frac{\partial (\Pi \boldsymbol{u})}{\partial
{\mathcal T}} =\eta\,\boldsymbol{u}\quad $ imply Eq. (\ref{system4}$^2$).

From relation
\begin{equation*}
\frac{\partial \Pi }{\partial \boldsymbol{u}}=\boldsymbol{j}^{\star
}\quad\Longrightarrow\quad \frac{\partial (\Pi \boldsymbol{u})}{\partial
\boldsymbol{u}}=\boldsymbol{u} \boldsymbol{j}^\star+\Pi\, \boldsymbol{I} ,
\end{equation*}
we obtain,
\begin{eqnarray*}
&&\func{div}\left( \frac{\partial (\Pi \boldsymbol{u})}{\partial \boldsymbol{u}%
}-\frac{\partial \Pi }{\partial q}\frac{\partial \boldsymbol{\Phi }}{%
\partial \boldsymbol{x}}-\frac{\partial \Pi }{\partial {\mathcal T} }\dfrac{%
\partial \boldsymbol{\Psi }}{\partial \boldsymbol{x}}\right) =\func{div}%
\left( \rho \,\boldsymbol{u}\,\boldsymbol{u}^{\star }\right) \boldsymbol{\ +}%
\frac{\partial \Pi }{\partial \boldsymbol{x}}\\
&&-\func{div}\left( \rho \,\frac{%
\partial \boldsymbol{\Phi }}{\partial \boldsymbol{x}}+\eta \,\frac{\partial
\boldsymbol{\Psi }}{\partial \boldsymbol{x}}\right) \\
&& = \func{div}\left( \rho \,\boldsymbol{u}\,\boldsymbol{u}^{\star }\right)
+\rho \,\frac{\partial \boldsymbol{\mu }}{\partial \boldsymbol{x}}+\eta \,%
\frac{\partial {\mathcal T}}{\partial \boldsymbol{x}}-\rho \,\frac{%
\partial \func{div}\boldsymbol{\Phi }}{\partial \boldsymbol{x}}-\eta \,\frac{%
\partial \func{div}\boldsymbol{\Psi }}{\partial \boldsymbol{x}}
\end{eqnarray*}
and consequently, the motion equation writes on form (\ref{system4}$^{3}$).

From relation
\begin{equation*}
\frac{\partial \Pi }{\partial \boldsymbol{\Phi }}=\boldsymbol{\beta }^{\star
}\quad\mathrm{and}\quad\frac{\partial (\Pi \boldsymbol{u)}}{\partial
\boldsymbol{\Phi }}+\frac{\partial \Pi }{\partial q}\,\dfrac{\partial
\boldsymbol{u}}{\partial \boldsymbol{x}}=\boldsymbol{u}\,\boldsymbol{\beta}%
^\star + \rho \,\frac{\partial \boldsymbol{u}}{\partial \boldsymbol{x}}=%
\frac{\partial (\rho \boldsymbol{u})}{\partial \boldsymbol{x}} \,,
\end{equation*}
we deduce Eq. (\ref{system4}$^{4}$).

From relation
\begin{equation*}
\frac{\partial \Pi }{\partial \boldsymbol{\Psi }}=\boldsymbol{\chi }^{\star
}\quad\mathrm{and}\quad\frac{\partial (\Pi \boldsymbol{u)}}{\partial
\boldsymbol{\Psi }}+ \frac{\partial \Pi }{\partial {\mathcal T}}\,\dfrac{\partial
\boldsymbol{u}}{\partial \boldsymbol{x}}=\boldsymbol{u}\,\boldsymbol{\chi}%
^\star + \eta \,\frac{\partial \boldsymbol{u}}{\partial \boldsymbol{x}}=%
\frac{\partial (\eta \boldsymbol{u})}{\partial \boldsymbol{x}}\,,
\end{equation*}
we deduce Eq. (\ref{system4}$^{5}$) and prove  System (\ref{system4}). \qquad\qquad\qquad\qquad\qquad\qquad$\square$

\end{document}